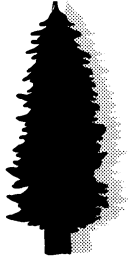

# ISSUES IN DYNAMICAL SUPERSYMMETRY BREAKING

MICHAEL DINE
Santa Cruz Institute for Particle Physics
University of California, Santa Cruz, CA 95064

ABSTRACT

Recent work has made clear that we have far more control over the dynamics of supersymmetric than non-supersymmetric theories. Here, I discuss some issues in dynamical supersymmetry breaking both in ordinary field theory and in string theory. In string theory, in particular, it is possible to show, in some circumstances, that stringy non-perturbative effects are *smaller* than effects visible in the low energy field theory. Observations of this sort suggest a general approach to string phenomenology.

## 1. Introduction

It is conventional to begin a talk on supersymmetry or superstring phenomenology with some remarks about supersymmetry's possible relevance to the hierarchy problem. One then usually says that while supersymmetry solves the problem of quadratic divergences, one has not yet understood how supersymmetry might be dynamically broken, and thus one does not understand the origin of the hierarchy.

The implication of these remarks is that somehow supersymmetric theories are particularly mysterious. However, as has been brought home most recently by the beautiful work of Seiberg and collaborators[1,2] and Seiberg and Witten,[3,4] supersymmetric theories are far more tractable than more conventional field theories. This fact has even been noted in the *New York Times Week In Review*, which devoted a recent article to the subject.[5] This article speaks of a "new mathematical tool...[whose] use is restricted, very restricted." In this talk, I hope to introduce you to this mysterious tool. While for mathematicians, this tool is truly new, for physicists the tool is neither so new nor so inaccessible. It consists largely in understanding the constraints imposed by supersymmetry on the low energy effective theory. In the past, it has been applied to problems of dynamical supersymmetry breaking, both in field theory and in string theory; this will be the subject of this talk. The dramatic new development is the application of this tool to explore a whole set of previously inaccessible problems in field theory, including aspects of quark confinement, weak and strong coupling duality, and certain aspects of the moduli spaces of strongly coupled supersymmetric theories, and, most startling in my view, the discovery of non-trivial duality relations between quantum field theories,[3,4] as well as non-trivial conformal field theories in four dimensions.[2] While I won't have time to even touch on most of these developments, the message which I hope to convey in this talk is that we do understand a great deal about these theories. Indeed, even in string theory, where we barely know what the theory is, we can say a great deal!

In light of the extraordinary understanding we have developed of supersymmetric field theories, it seems a good time to ask: If supersymmetry is realized in nature, how is it broken? The potential payoff of such an exercise is enormous:

1. One might predict features of the superparticle spectrum.
2. One will be led to predict new interactions (beyond those expected, say, in the MSSM) which might even be accessible experimentally.
3. One is likely to predict phenomena of relevance to cosmology (dark matter, moduli, etc.)
4. One might connect string theory to Nature.

I should stress from the start that there are problems one is unlikely to solve in the framework I will describe, such as the cosmological constant problem, and the question of what sort of principle might select a particular string vacuum.

In this talk, I will first introduce some basic aspects of supersymmetry, with emphasis on those features which make these theories so tractable. Two crucial features in this analysis will be the holomorphy of the gauge coupling function and the superpotential, and the enumeration of "flat directions" of the classical theory. I will then briefly discuss a class of globally supersymmetric models in which supersymmetry is broken with a stable ground state.[6] (These models, and their phenomenological prospects, will be discussed by Ann Nelson at this meeting). Finally, I will turn to string theory. I will discuss some of the flat directions of these theories and their significance. I will show that in many instances, any stringy non-perturbative effects are necessarily smaller than effects visible in the low energy theory, such as gluino condensation. Then I will turn to the problem of strong coupling in string theory, and argue that it may be possible to develop a phenomenology of strongly coupled string theory.[7] As applications of these ideas, I will consider the so-called "model-independent axion" of string theory, and show that it may provide a solution of the strong CP problem. I will also discuss string theory as a unified theory.

## 2. Supersymmetry and Supersymmetry Breaking

Two themes run through any discussion of supersymmetry and supersymmetry breaking:

   a. Flat directions: many supersymmetric models possess directions in field space in which the energy vanishes classically. The analysis of the dynamics in these directions is often very simple.
   b. Holomorphy in the fields: the low energy effective theory (in both global and local supersymmetry) is described by three functions of the chiral fields: the superpotential, $W$, the gauge coupling function, $f$, and the Kahler potential, $K$. The crucial point is that $W$ and $f$ are "holomorphic" (this means, in essence, analytic) functions of the chiral fields: $W = W(\phi), f = f(\phi)$. Their form is thus highly restricted. On the other hand, $K$ is essentially unrestricted, $K = K(\phi, \phi^\dagger)$.
   c. Holomorphy in the couplings: Much of the recent progress has followed from the realization that the couplings of a supersymmetric theory may themselves be viewed as expectation values of chiral fields. Thus the superpotential and gauge coupling functions are holomorphic functions of these parameters.[8,9,10] Thus $W$ and $f$ (but not $K$) are holomorphic functions of the couplings as well.

These ideas are illustrated by a simple model: an $SU(2)$ gauge theory with one "flavor," i.e., a doublet and antidoublet, $Q$ and $\bar{Q}$. We will first suppose that the classical superpotential vanishes This might occur, for example, if the underlying theory possessed a suitable discrete symmetry. In this case, the scalar potential is given by (here and in what follows we denote the scalar components of the chiral field by the same letter as the chiral field itself)

$$V = \tfrac{1}{2}g^2(Q^*T^aQ - \bar{Q}T^a\bar{Q}^*)^2. \tag{1}$$

If we take the expectation values of $Q$ and $\bar{Q}$

$$Q = \bar{Q} = \begin{pmatrix} v \\ 0 \end{pmatrix}, \tag{2}$$

then the scalar potential vanishes (for any choice of $v$). This means that classically the theory possesses a continuous set of physically inequivalent ground states. In each of these states, the gauge symmetry is completely broken and the spectrum consists of a massive gauge multiplet (the massive gauge bosons and their superpartners). In addition there is one massless state. The imaginary (CP-odd) part of this field is a conventional Goldstone boson; the real part arises because it costs no energy to change the expectation value $v$. This massless state can be written in terms of a gauge invariant field, $\Phi = \bar{Q}Q$. Expanding about the expectation values of the fields:

$$\Phi = \bar{Q}Q = v^2 + v(\delta\bar{Q} + \delta Q) + \ldots . \tag{3}$$

The effective coupling describing the low energy theory depends on $v$; it is basically $g^2(M_V)$, since the mass of the gauge bosons provides a lower cutoff on all Feynman integrals. By taking $v$ large we can make the coupling arbitrarily small.

This theory has one non-anomalous global symmetry. This symmetry is an "R" symmetry, which means that it does not commute with supersymmetry. The supercharges rotate by a phase under this transformation,

$$Q_\alpha \to e^{i\omega} Q_\alpha. \tag{4}$$

The scalar components of the fields $Q$ and $\bar{Q}$ transform as

$$Q \to e^{-i\omega}Q \qquad \bar{Q} \to e^{-i\omega}\bar{Q}. \tag{5}$$

The fermionic components have charge $-2$ under the symmetry.

It is crucial to what follows that, because supersymmetry is unbroken classically, the low energy effective action describing the light field will be supersymmetric, even if supersymmetry is spontaneously broken. Heuristically, this follows from the fact that supersymmetry breaking requires the existence of a massless fermion, in the case of global supersymmetry, and this must arise (at weak coupling) from the classically massless fields.[11] In the case of supergravity, it follows from the fact that, if supersymmetry is unbroken at tree level there is a massless spin-$\frac{3}{2}$ field. By general theorems, consistent coupling of such a field requires supersymmetry. As for any supersymmetric theory, then, the goal is to determine the superpotential, $W(\Phi)$, as well as the Kahler potential and the gauge coupling function, after integrating out the massive fields.

The effective lagrangian is constrained by the symmetries of the underlying theory, as well as by the holomorphy of the superpotential. In particular, it will respect the U(1)$_R$ of the underlying theory. Under an $R$ symmetry, the superpotential transforms as

$$W \to e^{2i\omega}W. \tag{6}$$

This fixes the forms of the superpotential *uniquely*:[12]

$$W_{np} = c\frac{\Lambda^5}{\bar{Q}Q} = c\frac{\Lambda^5}{\Phi}. \tag{7}$$

Here

$$\Lambda = e^{-8\pi^2/b_o g^2} \tag{8}$$

and $c$ is a constant of order one. $\Lambda$ is the non-perturbative scale of the SU(2) theory.

Even before attempting to evaluate the constant $c$, this is an extraordinary result. First, it implies that there is no renormalization of the superpotential in perturbation theory. Even more than that, it guarantees that non-perturbative corrections are not random, but are highly constrained. In fact, the constant $c$ can be obtained from a completely straightforward instanton computation, and shown to be non-zero.[13] This result means that supersymmetry is broken, but not in precisely the way one might like. The potential corresponding to Eq. (7) falls to zero at infinity – the region of weak coupling – while blowing up in the region of strong coupling, where one does not have control of the computation. This sort of runaway behavior is not uncommon, and will typify the situation in string theory.

One might hope to obtain a theory with a "nice vacuum" by adding a mass term, $m\bar{Q}Q$ so that the potential would rise at infinity. However, in this case, there are $N$ *supersymmetric* ground states, in accord with the Witten index theorem.[14] This can be seen by taking

$$W = mQ\bar{Q} + W_{np} \tag{9}$$

and solving

$$\frac{\partial W}{\partial Q} = \frac{\partial W}{\partial \bar{Q}} = 0.$$

What sorts of models exhibit supersymmetry breaking with a stable ground state? The following two conditions appear to be sufficient:
  *a*. The classical theory should possess no flat directions.
  *b*. The theory should possess a global symmetry, which is spontaneously broken.

It is easy to see why one expects supersymmetry breaking under these circumstances: Suppose supersymmetry were unbroken. Then, the Goldstone boson would be accompanied by a (scalar) superpartner. However, because there is no potential for the Goldstone boson, there is no potential for the superpartner. This implies the existence of a flat direction, which violates the original assumption (unless the partner is itself a Goldstone boson, or the flat direction is compact). This argument is rather heuristic, but in many cases one can calculate the non-perturbative potential, and show that this "theorem" is satisfied. The simplest known such model has gauge group SU(3) × SU(2), and a set of chiral fields simi-lar to that of one generation of the standard model, without the positron:[15]

$$Q\ (3,2) \quad \bar{u}\ (\bar{3},1) \quad \bar{d}\ (\bar{3},1) \quad L\ (1,2). \tag{10}$$

At the classical level, one also adds the most general superpotential allowed by the gauge symmetries:

$$W = \lambda Q L \bar{d}. \tag{11}$$

It is not hard to check that this model has no flat directions. It does possess two global U(1) symmetries. If $\lambda$ and the SU(2) gauge coupling are small, then the theory reduces to SU(3) gauge theory with two flavors, a theory in which instantons generate a superpotential. Adding this to the classical superpotential, one finds that the true minimum has broken supersymmetry. The vacuum energy, spectrum, and other features of the theory can then be computed systematically.[15,16]

Until recently, only a few such models were known. In the last few months, however, many additional models exhibiting dynamical supersymmetry breaking have been discovered.

    *a.* Along the same lines as above, Poppitz and Randall suggested looking at models with U(1)'s. A. Nelson and I have indeed found a large set of such models, as well as models with other group structures. These include models with flat directions lifted by renormalizable and non-renormalizable operators.[17]

    *b.* Seiberg, Intriligator and Leigh have developed a much more comprehensive understanding of the dynamics of such theories. This includes, in many cases, an understanding of the strong coupling, $v \to 0$, region of these theories. This permits analysis of a much broader class of models.[1] Exploiting this improved understanding of dynamics, Intriligator, Shenker and Seiberg have discovered a very simple model, an SU(2) theory with a chiral field in the isospin-3/2 representation in which supersymmetry is broken. The model has many features which are qualitatively different than earlier known models. For example, in the limit that the superpotential is turned off, there are exact flat directions non-perturbatively.[18]

## 3. Model Building

Roughly speaking, we have seen that there are three classes of supersymmetric models. There are some in which supersymmetry is unbroken, even nonperturbatively. There are some in which supersymmetry is broken, with a "nice," stable vacuum state. And there are some in which the classical theory exhibits flat directions, which are lifted non-perturbatively, but without yielding a stable vacuum, at least at weak coupling or before including non-renormalizable operators.

String models inevitably fall into this latter class. In this section, we will focus on trying to build realistic models of particle physics using models of the second type, with broken supersymmetry and stable ground states.

The first question one must address, in trying to build such models, is that of the scale of supersymmetry breaking. There are many possibilities which one might consider. One is to use one of these models as the "hidden sector" of a conventional supergravity model, replacing, e.g., the "Polonyi field." In such a model, the scale of supersymmetry breaking will be the scale, $M_I = \sqrt{m_w M_P}$. In trying to build such models, however, one immediately encounters several problems. The most severe is that of obtaining a sufficiently large gaugino mass and "$\mu$" term. Gaugino masses must arise through operators of the form

$$\int d^2\theta W_\alpha^2 f(\Phi). \tag{12}$$

Here $f(\Phi)$ is a holomorphic function of the chiral fields. However, in the models with DSB the lowest dimension, gauge-invariant functions which one can write are of dimension three, i.e.,

$$f(\Phi) \to \frac{\Phi^3}{M_p^2}. \tag{13}$$

This leads to a huge suppression of these masses. Some loopholes to this argument exist, but it is not at all clear that gaugino masses can be obtained consistent with the current severe experimental constraints which we have heard about at this meeting.[19]

Similar remarks apply to the $\mu$ parameter. Such a term can arise from terms in the Kahler potential of the form

$$g(\Phi^\dagger)H_1 H_2. \tag{14}$$

where $\Phi$ represent, once more, hidden sector fields. However, in order to obtain a term of the correct order of magnitude, $g$ must have dimension one, and again there are no such gauge invariant operators.

If one could think of ways around these difficulties, these models would have certain advantages over conventional hidden sector models, apart from the dynamical origin of supersymmetry breaking. For example, because they do not possess even approximate flat directions, they do not suffer from the severe cosmological problems of more conventional models.[20]

An alternative possibility is that supersymmetry is broken at comparatively low energies, within a few decades of the weak scale. In models which have been constructed up to now, the basic strategy has been, again, to break supersymmetry in a "hidden" sector. However, now supersymmetry breaking is fed down to the partners of ordinary quarks and leptons through gauge interactions. Various obstacles to constructing such models have been pointed out in the past.[15] The most serious of these are problems of axions and of asymptotic freedom. The former problem has been solved in the past year by Bagger, Poppitz and Randall.[16] This solution, along with a solution of the latter problem, has been incorporated in the models of ref. 6. These models have been discussed in some detail by Ann Nelson at this meeting. Let me just mention their principle virtues:

    *a*. Calculability: all of the soft breakings are calculable. As a result, the models make detailed predictions for the superparticle spectrum.

    *b*. Because squark and slepton masses arise through gauge interactions, they are functions only of gauge quantum numbers, to first approximation. As a result, flavor changing processes are more than adequately suppressed.

    *c*. Finally, there is rich new physics "nearby." In all such models, first of all, fields beyond those of the MSSM are required to obtain $SU(2) \times U(1)$ breaking. In particular, gauge singlet fields seem to be essential. Second, the interactions responsible for supersymmetry breaking themselves lie at multi-TeV scales, and may have additional implications for low energy physics.

## 4. Strongly Coupled String Theory

As we have explained, two themes run through any discussion of dynamical supersymmetry breaking: flat directions and holomorphy. These permit one to make powerful statements about the non-perturbative dynamics of the theory, even in some cases at strong coupling. String theories are an obvious arena in which to explore these ideas. String theories are notorious for flat directions. Moreover, in string theory not only is non-perturbative physics difficult – we do not even possess a non-perturbative definition of the theory! As we will see, the methods we have described earlier permit one to make powerful statements about these theories. They lead, for example, to a two line argument for the finiteness of string theory.[21] More strikingly, they permit us in many instances to place strong bounds on the size of any inherently stringy non-perturbative effects. Perhaps most important of all, they suggest how, even if string theory is strongly coupled, it may yet make contact with reality. The most simple and dramatic prediction of this framework is that – even at strong coupling – the theory, at low energies, should look like a supersymmetric theory with small, explicit soft breakings.

It is perhaps worth taking a moment to recall some of the promising and not so promising features of string theory. Perhaps the most compelling feature of string theory is that it is the only known quantum mechanically consistent theory of gravity and gauge interactions. But there are added bonuses. The theory possesses classical ground states in which:

1. The universe is four dimensional.
2. The gauge group is close to that of the standard model.
3. There are repetitive chiral generations of quarks and leptons.
4. Many of these ground states are supersymmetric.
5. The couplings of the fields are calculable, in principle.

There are, however, a number of difficulties which stand in the way of developing a convincing string phenomenology, and which cast doubt on the theory's ultimate truth:

1. There is a huge multiplicity of vacua, involving both discrete choices (e.g., the number of generations) and continuous parameters (the "moduli").
2. To date, the theory has provided no insight into the cosmological constant problem. Perhaps the theory will yet produce, in some circumstances, miraculous cancelations between high and low energy physics. Alternatively, it may lead us to different ways of thinking about this problem.
3. If the theory does describe nature, one can argue (though not quite prove) that it must be strongly coupled.[22] This is troubling, for several reasons. First, our experience with strongly coupled field theory indicates that the spectrum at strong coupling may be completely different than that at weak coupling. This would be disappointing, since it is precisely the spectrum of the weakly coupled theory which is what is appealing about string theory. Second, even if this is not the case, since we know almost nothing of the non-perturbative theory, it is hard to see how we might make contact with nature. Even when we have a non-perturbative formulation, strong coupling computations are not likely to be feasible. Third, the forces we know in nature are all weakly coupled. Fourth (and related to the first point), while we have argued that one of the beauties of the theory is that it produces low supersymmetry at low energies, if the theory is strongly coupled, it is not at all clear why this feature should survive into the true ground state.

We have hinted above that the constraints imposed by supersymmetry itself may provide a way out of these dilemmas. To understand why holomorphy is such a powerful tool in dealing with string theory, it is necessary remember that the dimensionless coupling constant of string theory may itself be thought of as the expectation value of a dynamical scalar field. In four dimensional string models which preserve $N = 1$ supersymmetry, this field is a component of a chiral superfield, usually denoted by the letter $S$:

$$S = \frac{8\pi^2}{g^2} + D + ia + \dots . \qquad (15)$$

Here $D$ represents the fluctuating part of the dilaton field. $a$ is an axion, which couples universally to the $F\tilde{F}$ of each gauge group. (The decay constant of this axion is of order the $M_P$.) $g$ here represents the gauge coupling at the string scale. In the real world, then, one might imagine that $S$ is a number of order 300 or so.

In string perturbation theory, it is not hard to show that the axion is truly an axion, i.e., the theory is symmetric under shifts[23]

$$a \to a + \delta \tag{16}$$

or

$$S \to S + i\delta. \tag{17}$$

On the other hand, the superpotential must be a holomorphic function of $S$. This means that the superpotential is independent of $S$, i.e., the superpotential is not renormalized![21]

Since the superpotential is independent of $S$, we have also learned that the theory possesses, classically and in perturbation theory, an exact flat direction; there are a continuum of physically inequivalent ground states, characterized by different values of the coupling. As we have mentioned, in string theory the degeneracy is typically much larger, extending to other "moduli." In what follows, we will focus, for simplicity, on the dilaton; all of our remarks are readily extended to these other moduli.

Just as in field theory, we want to investigate what happens beyond perturbation theory. In particular, as in field theory, we expect that the Peccei-Quinn symmetry is broken, and that the non-renormalization theorem may break down. All existing work on this problem involves examining the low energy effective theory for effects which might break supersymmetry. One sort of breakdown which has been widely discussed in the literature is "gluino condensation in a hidden sector.[24] The point here is that if one has a pure gauge sector of the theory (no chiral matter) with gauge group, say, SU(N), then condensation of the gluinos of this sector gives rise to a superpotential which behaves as

$$W(S) \sim e^{-S/N}. \tag{18}$$

Before dealing with any phenomenological issues, a natural question to ask is whether even for weak coupling (large $S$), gluino condensation (we use this term as a code word for any supersymmetry-breaking effects visible in the low energy effective theory) is the largest non-perturbative effect. After all, who says that there can't be much larger effects (behaving, say, as $e^{-1/g}$), arising from integrating out massive string modes?

It turns out that in many cases, one can argue that gluino condensation is larger than any possible non-perturbative effect. As usual, the trick involves exploiting both holomorphy and symmetries. The idea is that while the Peccei-Quinn symmetry is broken non-perturbatively, there are often discrete gauge symmetries which can play a similar role. Most powerful for this purpose are discrete $R$ symmetries. Many string vacua exhibit such symmetries. For our purposes, there are only three important facts one needs to know. First, as for most discrete symmetries of string theory, these symmetries can be thought of as gauge symmetries of some kind, and are thus expected to survive beyond perturbation theory. Second, thought of as a function of chiral fields, the superpotential in general must transform by a phase under such symmetries. Finally, in many instances, the axion transforms under these symmetries,[25] e.g.,

$$a \to a + \frac{2\pi k}{N} \qquad (19)$$

in the case of a $Z_N$ symmetry.

The consequences of these symmetries are again quite dramatic. If the axion transforms as in Eq. (19), for example, then any term in the superpotential involving $S$ alone must be of the form

$$W = e^{-rS}$$

where $r$ is a rational number, related to $k$ and to the transformation properties of the superpotential (e.g., if $k = 1$ and $W \to e^{2\pi i/N} W$, $r = 1$). More generally, $S$ can only appear in the superpotential through exponentials of this form. In many instances, $r$ is such that these terms are smaller than those which arise from gluino condensation.

One can actually give a somewhat shakier argument that the non-perturbative theory must respect a $2\pi$ periodicity of the axion.[7] If this is really true, only terms of the form $e^{-nS}$ are permitted in the superpotential. In any case, one can often reliably assert that the low energy supersymmetry breaking effects are the largest effects at weak string coupling.

Before exploring any particular mechanism for supersymmetry breaking, however, we can see that we face a fundamental difficulty. Whatever non-perturbative effects may lift the $S$ flat direction, they necessarily tend to zero at very weak coupling.[22] So the $S$ potential always resembles that of the models we discussed in section 2. If string theory does describe nature, it is necessary that the potential for $S$ turns over at some relatively small value of $S$. By definition, this is strong coupling.[*] At first sight, one might despair, since it is not clear whether any of the good features of string theory should survive to strong coupling, and in any case there is no foreseeable prospect for doing strong coupling string computations.

---

[*] The only serious suggestion for avoiding this problem is the "racetrack approach" of ref. 26. To date, however, it is probably fair to say that no satisfactory model of this type has been found.

However, our discussion of constraints on non-perturbative stringy effects suggests that the problem may not be nearly so severe, and offers a path to a phenomenology of strongly coupled strings. There is good reason to believe that perturbation theory in string theory is far less convergent than field theory perturbation theory. Thus one can imagine that, for $S \sim 300$, say, the string perturbation expansion is not valid, and that there are $\mathcal{O}(1)$ corrections to tree level results. Still, $e^{-S}$ would be an incredibly small number. So corrections to the superpotential and the gauge coupling function would be very tiny. On the other hand, the Kahler potential would receive order one corrections. As a result, one could imagine that the true vacuum of string theory is obtained from a classical ground state by varying $S$. Gluino condensation would give a superpotential for $S$, and *corrections to the Kahler potential would stabilize the potential* at $S \sim 300$. It is not hard to write down functional forms for $K$ which would do the job. In this situation, one can argue that

1. The light spectrum in the true vacuum is the same as at weak coupling.
2. The low energy effective theory is a theory which is approximately supersymmetric with small, explicit soft breakings.
3. The symmetries of the theory are the same as at weak coupling (there are no phase transitions).
4. The superpotential of the matter fields is identical to that of the free theory. This does not mean that Yukawa couplings are unchanged, since these depend on the Kahler potential. However, typically some relations among Yukawa couplings are unchanged.
5. The gauge couplings are unified at the string scale (in the sense of ref. 8).

All of these statements are consequences of holomorphy and discrete symmetries, i.e., the fact that $e^{-rS}$ is numerically very small. (More detailed arguments can be found in ref. 7.) This is not, perhaps, all one might have wished for. All quantities which depend on the Kahler potential are unknown in this approach, so one cannot determine the pattern of soft breakings. For example, the schemes suggested by Ibanez et al. and by Kaplunovsky and Louis for obtaining squark degeneracy cannot be implemented in this approach, since they depend critically on properties of the lowest order Kahler potential.[27] Still, these observations suggest an approach to string phenomenology not too different than that pursued by many authors at the present time: search for string models with a desirable set of gauge symmetries and light fields; examine symmetries of the model; compute the tree level superpotential. Indeed, this view of strong coupling provides the only clear rationalization for such an approach.

The idea of stabilization of the dilaton through corrections to the Kahler po-tential suggests ways of thinking about other problems which have not been considered before. Consider the strong CP problem. It is often said that the model-independent axion of string theory cannot be the QCD axion, because it will gain a substantial mass through hidden sector interactions. But this is not always true. Consider, for example, a model with a $Z_5$ symmetry (as in the simplest Calabi-Yau model in ref. 23). Suppose that supersymmetry is broken through a single gluino condensate. At the level of renormalizable couplings, the hidden sector possesses an anomalous U(1) symmetry. So, ignoring non-renormalizable couplings, the axion would not gain mass. Because of the $Z_5$ symmetry, the leading operators which could break this U(1) symmetry (not even worrying about supersymmetry) would be of the form $(\lambda\lambda)^5$, i.e., they would be operators of dimension 15! As long as the hidden sector scale was not extremely large, this would more than adequately suppress the axion potential.

A second issue often discussed is the possibility that string theory could behave like a GUT. It is argued that this would help with the problem of unification of couplings. Without asking how serious this problem really is (recent analyses discussed at this meeting suggest that the problem is perhaps not much more serious in string theory than in supersymmetric GUTS[28]) we can ask whether this idea is particularly promising. Suppose that one has a string model with some set of massless adjoint fields. One requires that these fields obtain a vev of order $10^{16}$ GeV, i.e., quite large. At the same time, these vevs should not break supersymmetry by too large an amount. Again, the superpotential will not be corrected to any order in perturbation theory; non-perturbative corrections are almost surely too small to be relevant. So how is such a vev to arise? If the adjoint vev arises through the tree level potential, then the natural value for the adjoint vev is the string scale. One could hope that somehow a small number appears fortuitously. But this would simply mean that there exists another string vacuum, in which some states, with masses parametrically of order the string scale, are fortuitously a couple of orders of magnitude lighter. But then it was not unification, but some numerical accident, which explained the meeting of the couplings.

This suggests that any meaningful unified scenario requires the existence of flat directions for the adjoint field. For example, due to discrete symmetries, there might be no potential at all for this field. But in this case, there is a new potential difficulty: it is necessary that the octet and triplet components of the octet gain large mass in the flat direction.[29] In simple SU(5) models, this can arise, though from a field-theoretic perspective it looks finely tuned. In O(10) models, one can arrange this, though it is not clear that the required particle content (three adjoints, for example,[30]) arises in string models.[31] Of course, even if one does solve this problem, one still needs to understand the doublet/triplet splitting problem for the ordinary Higgs. While one can probably propose solutions to all of these difficulties, they are likely to be quite complicated.

## 5. Conclusions

Theories with low energy supersymmetry are astoundingly tractable. The examples discussed here represent only a small sample of the questions which have been attacked by these methods. It is almost certain that they further shed light on many issues, both in string theory and field theory. Perhaps they will yet provide some insight into some of the hardest questions we face, such as the problem of the cosmological constant and the existence (which we have here only assumed) of a strongly coupled string ground state.